\documentclass[preprint,amsmath,amssymb,aps,pre,floatfix]{revtex4-1}

\usepackage{graphicx}
\usepackage{dcolumn}
\usepackage[hidelinks]{hyperref}
\usepackage{bm}
\usepackage[mathlines]{lineno}

\hypersetup{
    colorlinks,
    linkcolor={blue},
    citecolor={blue},
    urlcolor={blue}
}

\begin{document}\

\title{Experimental investigation of vertical turbulent transport of a passive scalar in a boundary layer: statistics and visibility graph analysis}

\author{G. Iacobello}
	\email{giovanni.iacobello@polito.it}
	\affiliation{Department of Mechanical and Aerospace Engineering, Politecnico di Torino, Turin, Italy}
\author{M. Marro}
	\affiliation{Laboratoire de M\'ecanique des Fluides et d'Acoustique, University of Lyon, CNRS UMR 5509, {\'E}cole Centrale de Lyon, INSA Lyon, {\'E}cully, France}
\author{L. Ridolfi}
	\affiliation{Department of Environmental, Land and Infrastructure Engineering, Politecnico di Torino, Turin, Italy}	
\author{P. Salizzoni}
	\affiliation{Laboratoire de M\'ecanique des Fluides et d'Acoustique, University of Lyon, CNRS UMR 5509, {\'E}cole Centrale de Lyon, INSA Lyon, {\'E}cully, France}
\author{S. Scarsoglio}
	\affiliation{Department of Mechanical and Aerospace Engineering, Politecnico di Torino, Turin, Italy}

\begin{abstract}
		
		The dynamics of a passive scalar plume in a turbulent boundary layer is experimentally investigated via vertical turbulent transport time-series. Experimental data are acquired in a rough-wall turbulent boundary layer that develops in a recirculating wind tunnel set-up. Two source sizes in an elevated position are considered in order to investigate the influence of the emission conditions on the plume dynamics. The analysis is focused on the effects of the meandering motion and the relative dispersion of the plume with respect to its center of mass. First, classical statistics are investigated. We found that (in accordance with previous studies) the meandering motion is the main responsible for differences in the variance and intermittency, as well as the kurtosis and power spectral density, between the two source sizes. On the contrary, the mean and the skewness are slightly affected by the emission conditions. With the aim to characterize the temporal structure of the turbulent transport series, the visibility algorithm is exploited to carry out a complex network-based analysis. In particular, two network metrics -- the average peak occurrence and the assortativity coefficient -- are analysed, as they are able to capture the temporal occurrence of extreme events and their relative intensity in the series. The effects of the meandering motion and the relative dispersion of the plume are discussed in the view of the network metrics, revealing that a stronger meandering motion is associated with higher values of both the average peak occurrence and the assortativity coefficient. The network-based analysis advances the level of information of classical statistics, by characterizing the impact of the emission conditions on the temporal structure of the signals in terms of extreme events (namely, peaks and pits) and their relative intensity. In this way, complex networks provide -- through the evaluation of network metrics -- an effective tool for time-series analysis of experimental data.

\end{abstract}

\maketitle

\section{Introduction}
		
		The release in the atmospheric boundary layer of flammable or toxic substances, as well as the dispersion of pollutants, need to be carefully addressed due to their significant environmental and health impact. To this end, different numerical and experimental strategies have been adopted so far \cite{arya1999air}, in order to investigate the relation between turbulence dynamics, release conditions and one-point probability density functions (PDFs) of the pollutant concentration. In particular, several works investigated the relation between statistical moments, in order to infer the corresponding PDF \cite{chatwin1990simple, mole1995relationships, yee1997comments, schopflocher2005relationship, vinkovic2006large, bisignano2017evaluation, marro2018dispersion}. For example, \textit{Chatwin and Sullivan} \cite{chatwin1990simple} investigated the relation between mean and standard deviation of a passive scalar in a shear flow, while \textit{Mole and Clarke} \cite{mole1995relationships} found a functional dependency between the third- and fourth-order moments. In other works, instead, the form of the PDF of the concentration was directly investigated (see, among others, \cite{yee1997simple,villermaux2002mixing}), with implications on the passive scalar modelling.
		
		Although several set-up configurations can be adopted, the release of a contaminant from a point source in a turbulent boundary layer provides an accurate representation of a typical plume dispersion process in the atmospheric boundary layer \cite{arya1999air}. From a modelling perspective, \textit{Gifford} \cite{gifford1959statistical} described the dynamics of a plume continuously emitted by a point-source as mainly governed by two mechanisms: the meandering motion of the instantaneous center of mass of the plume, and the relative dispersion -- i.e., spreading -- of the plume with respect to its center of mass. The plume meandering motion is due to turbulent length scales larger than the plume size. Namely, only the large scale eddies of the turbulent flow are able to make the plume meander in space. The two main parameters that affect the meandering of a developing plume are the size and the distance of the source from the ground. For an elevated source, plumes emitted by a smaller source size are affected by a wider range of turbulent length scales (thus implying a stronger meandering motion) with respect to a plume emitted by a larger source size. As shown in previous studies \cite{fackrell1982concentration, nironi2015dispersion}, small variations of the source size significantly affect the role of meandering in the plume dispersion, by inducing variations in the (one-point) concentration statistics up to streamwise distances of the order of a hundred times the source size. For a ground-level source, instead, the plume dynamics is slightly affected by the source size \cite{fackrell1982concentration, nironi2015dispersion, talluru2018local} because close to the ground turbulent length scales are typically of the same order of magnitude of the source size. The effect of turbulent eddies nominally smaller than the plume size, instead, is to contribute to the local mixing of the concentration field, thus promoting the relative dispersion of the plume. 
		
		In this work, we address the same issue studied by \textit{Fackrell and Robins} \cite{fackrell1982concentration} and \textit{Nironi et al.} \cite{nironi2015dispersion} to investigate the spatio-temporal development of a passive scalar plume, emitted from an elevated point-source of varying size. In particular, experimental measurements of velocity and passive scalar concentration are performed in a turbulent boundary layer over a rough-wall, which is intended to represent the dispersion process of a passive scalar in the atmospheric boundary layer. The present analysis highlights the effects of the meandering and the relative dispersion on the wall-normal turbulent transport of the passive scalar. The turbulent transport is here investigated as it plays a key role into the interplay between the turbulence velocity field and the concentration of passive scalar. In fact, due to the presence of the ground, the extent to which the passive scalar is transported along the wall-normal direction is a fundamental aspect for the dispersion characterization. Specifically, the role of the different source size on the plume dynamics, as well as the plume spatial evolution along the streamwise and wall-normal coordinates, is emphasized throughout the study. 
		
		In order to investigate the plume dynamics, two approaches are carried out: (i) we obtain statistics of wall-normal turbulent transport, thus enriching the benchmark for a dispersing plume in a rough-wall set-up \cite{fackrell1982concentration, nironi2015dispersion}; (ii) a complex network-based analysis is performed, in order to advance the level of information of classical statistical tools, thus revealing non-trivial insights into the temporal structure of the signals. Although different research fields have taken advantage of network science during last decades (e.g., social, biological, or technological networks \cite{newman2018networks}), only recently complex networks have emerged as an effective framework also to study fluid flows. The main applications of network science to fluid flows involve the study of two-phase flows \cite{gao2009flow, gao_ijbc_2017}, turbulent jets \cite{charakopoulos2014application, murugesan2019complex}, isotropic and wall-bounded turbulence \cite{taira2016network, iacobello2018visibility, iacobello2018spatial}, reacting flows \cite{unni2018emergence, murugesan2015combustion}, Lagrangian mixing \cite{iacobello_2019,padberg2017network} and geophysical flows \cite{ser2015flow, charakopoulos2018dynamics}. Among several techniques that have been developed so far to study time-series by means of complex networks \cite{zou2018complex}, the visibility graph approach \cite{Lacasa2008} was here adopted since it is a simple but powerful tool to extract non-trivial insights into the non-linear process from which the time-series are obtained \cite{zou2018complex}.

		The paper is organized as follows. Section \ref{sec:methods} includes the description of experimental set-up and the measurement techniques (Section \ref{subsec:setup}), as well as the data pre-processing (Section \ref{subsec:preproc}). The statistical analysis of the wall-normal turbulent transport is reported in Section \ref{sec:statistics}. Typical statistical quantities are evaluated, such as the first four moments (i.e., the mean value, the standard deviation, the skewness and kurtosis), the power spectral density and the intermittency factor. The complex network analysis is shown in Section \ref{sec:visib}: the main concepts of complex network and visibility graph are outlined in Section \ref{subsec:net_def}, while the metrics definition and interpretation is discussed in Section \ref{subsec:net_metr}. Two network-metrics are investigated -- i.e., the average peak occurrence and the assortativity coefficient \cite{iacobello2018visibility} -- with the aim to characterize the temporal structure of the turbulent transport time-series. The average peak occurrence and the assortativity coefficient are here selected as they are able to highlight the temporal structure of extreme events (i.e., peaks and pits) and their relative intensity in a time-series. The results of the visibility-network approach are shown in Section \ref{subsec:metr_analysis}, while the conclusions are drawn in Section \ref{sec:conclus}.

\section{Methods}\label{sec:methods}

	\subsection{Experiment set-up and measurements}\label{subsec:setup}

	\begin{figure*}[t]
		\centering
		\includegraphics[width=.85\linewidth]{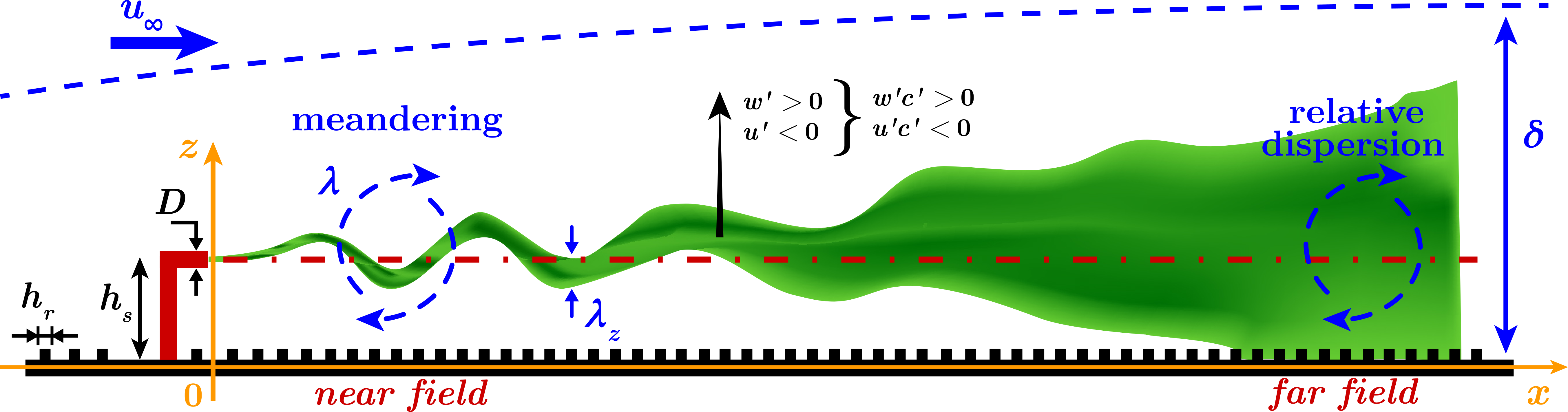}
		\caption{Sketch of the TBL set-up; the plume is illustrated in green, while the horizontal dash-dot line refers to the source axis. The symbols are defined in the main text. Two large-scale eddies of (Eulerian) characteristic size $\lambda$ are also depicted as rotating arrows.}
		\label{fig:metod}
	\end{figure*}

		A neutrally-stratified atmospheric turbulent boundary layer (TBL) was generated in a recirculating wind tunnel of the Laboratoire de M{\'e}canique des Fluides et d'Acoustique at the {\'E}cole Centrale de Lyon, in France. The set-up and the measurement tools are the same as that adopted by \textit{Nironi et al.}\cite{nironi2015dispersion} (see Appendix \ref{app:DET}). However, measurements were performed in a wind tunnel which smaller than that used by \textit{Nironi et al.} \cite{nironi2015dispersion}, with a working section that is $9$ m long, $1$ m wide, and $0.7$ m high. A row of Irwin spires \cite{irwin1981design} were placed at the beginning of the test section, while cubic roughness elements with size $h_r=0.02$ m were uniformly displaced on the floor. As a result, a TBL of free-stream velocity $u_\infty=4.94$ $\textrm{m/s}$ and thickness $\delta=0.314$ m was generated, with $\delta$ evaluated as the wall-normal coordinate where the mean velocity $\overline{u}=0.95 u_\infty$ (see a sketch of the set-up in Fig. \ref{fig:metod}). The Reynolds number of the experiment was evaluated as ${Re}_\delta=\delta u_\infty/\nu\approx1.034\times10^5$ ($\nu=1.5\times10^{-5}$ $\textrm{m}^2/\textrm{s}$ is the kinematic viscosity of air), which guarantees a well-developed rough turbulent flow \cite{jimenez2004turbulent}. In this work, the streamwise, spanwise and wall-normal directions are indicated as $(x,y,z)$, respectively, and the origin of the axes is at the wall in correspondence to the outlet section of the source (see the sketch in Fig. \ref{fig:metod}).
		\\A mixture of air and a passive scalar was continuously ejected from a metallic L-shaped tube. Due to its density similar to air, Ethane ($\textrm{C}_2\textrm{H}_6$) was used as a passive tracer. The passive scalar source was located at a streamwise distance from the beginning of the working section $x_s/\delta\approx 17.5$ and at a wall-normal height $h_s/\delta\approx 0.24$. Two internal diameter configurations were considered (see Fig. \ref{fig:metod}): $D=0.003$ m (i.e., $D/\delta\approx9.55\times10^{-3}$) and $D=0.006$ m (i.e., $D/\delta\approx1.91\times10^{-2}$). In the following, these two configurations are referred to as D3 and D6, respectively. The Ethane-air mixture does not substantially introduce or subtract momentum from the flow field at the source. This condition is referred to as \textit{isokinetic} \cite{nironi2015dispersion, talluru2018local}, namely the source velocity, $u_s$, of the mixture equals the local mean velocity, $\overline{u}$, at the source height, $u_s\equiv \overline{u}(z = h_s)\approx 3.37$ $\textrm{m/s}$. In order to have isokinetic conditions, the total mass flow rate, $M_t=\rho u_s \pi D^2/4$, was imposed as $M_t/\rho\approx 86$ $\textrm{l}/\textrm{h}$ for D3 and $M_t/\rho\approx 344$ $\textrm{l}/\textrm{h}$ for D6, where $\rho=\rho_{air}=\rho_{C2H6}$ is the density of the air-Ethane mixture. Furthermore, to consider the recirculation of Ethane-air in the wind tunnel, the background concentration (which increases linearly with time) was subtracted from the recorded time-series.

	The streamwise and vertical velocity time-series, $u$ and $w$, were acquired by means of a X-probe hot-wire anemometer (HWA), while concentration time-series, $c$, were recorded with a fast Flame Ionization Detector (FID) \cite{fackrell1980flame} (for further details on the instruments see Appendix \ref{app:DET}). The acquisition time was set equal to $T=180$ s while the number of recorded data is $N_T=1.8\times10^5$. Measurements were performed at different locations along the three Cartesian directions, $(x,y,z)$. Specifically, data were recorded at $x/\delta=\{0.325, 0.650, 1.30, 2.60, 3.90\}$ in the streamwise direction. For each $x/\delta$ location, one-point measurements were taken along the vertical (i.e., at fixed $y/\delta$) and transversal (i.e., at fixed $z/\delta$) directions. Transversal profiles of concentration and velocity are obtained at $z=h_s$, and at spanwise locations ranging in the interval $y/\delta=\left[-0.6,0.6\right]$. On the other hand, vertical profiles are obtained at $y/\delta=0$ at various wall-normal locations ranging in the interval $z/\delta=\left[0.096,0.828\right]$ (the limits depend on the estimated size of the plume at a given $x/\delta$). Due to the crucial role played by the wall-normal direction that represents the direction of spatial inhomogeneity of the flow, here we focus on measurements taken at $y/\delta=0$, namely in the $(x,z)$-plane normal to the wall and passing through the source axis (Fig. \ref{fig:metod}). We refer to the \textit{near field} and the \textit{far field} as the streamwise locations closest and farthest from the source, respectively. Therefore, the near and far fields correspond to locations $x/\delta=0.325$ and $x/\delta=3.90$, respectively, while $x/\delta=1.30$ is considered as an intermediate location.

	\subsection{Data pre-processing}\label{subsec:preproc}	
	
		Each time-series was normalized by a reference value, which is $u_\infty$ for the velocity components (measured in $\textrm{m/s}$) and $\Delta c=M_e/\left(\rho u_\infty \delta^2\right)$ for the passive scalar concentration (measured in ppm), where $M_e$ is the mass flow rate of Ethane. Therefore, in this work we indicate as $c$, $u$ and $w$ the normalized concentration, streamwise and wall normal velocity, respectively. To take into account the presence of random instrumental noise on $c$ -- that produces negative concentration values -- we preprocessed the concentration data as

		\begin{equation}\label{eq:c_eps}
			c(x,z;t_i)=0,\quad\textrm{if}\quad c(x,z;t_i)<\epsilon,
		\end{equation}

		\noindent where $\epsilon=\left|\min_{x,z}\left[\min_i{\left[c(x,z;t_i)\right]}\right]\right|$ is the absolute value of the minimum amplitude of all concentration series for a given $D$. In other words, we set equal to zero all concentration values that are smaller (in modulus) than the maximum amplitude of negative values in the series. This pre-processing operation is reasonably valid as the values of $\epsilon$ are two orders of magnitude lower than the average concentration values, and three orders of magnitude lower than the maximum $c$ values. Furthermore, since in this work we focus on the vertical passive scalar flux, $w'c'$, the Reynolds decomposition was performed for velocity and concentration time-series as $w'=w-\overline{w}$ and $c'=c-\overline{c}$, where $\overline{w}$ and $\overline{c}$ are the time-averages of $w$ and $c$, respectively. 		
		
		Finally, we estimated the vertical position, $h_s^*$, of the actual axis of the plume, for both D3 and D6, as the $z$ coordinate of maximum $\overline{c}(z)$ value (see Appendix \ref{app:PAE} for more details). In fact, since the plume develops in a turbulent boundary layer, it is affected by the mean shear and by the source wake. While the latter is mainly present very close to the source, the mean shear acts at any streamwise location and tends to tilt the plume axis towards the wall. As a consequence, the wall-normal coordinate of the plume axis, $h_s^*$, along $z/\delta$ is not exactly at $z=h_s$, but it decreases downstream from the source. Although the values of $h_s^*$ for D3 and D6 should be different, this is true only in the near field, i.e. where the differences between the plumes emitted by D3 and D6 are the strongest.

		\begin{figure*}[t]
			\centering
			\includegraphics[width=\linewidth]{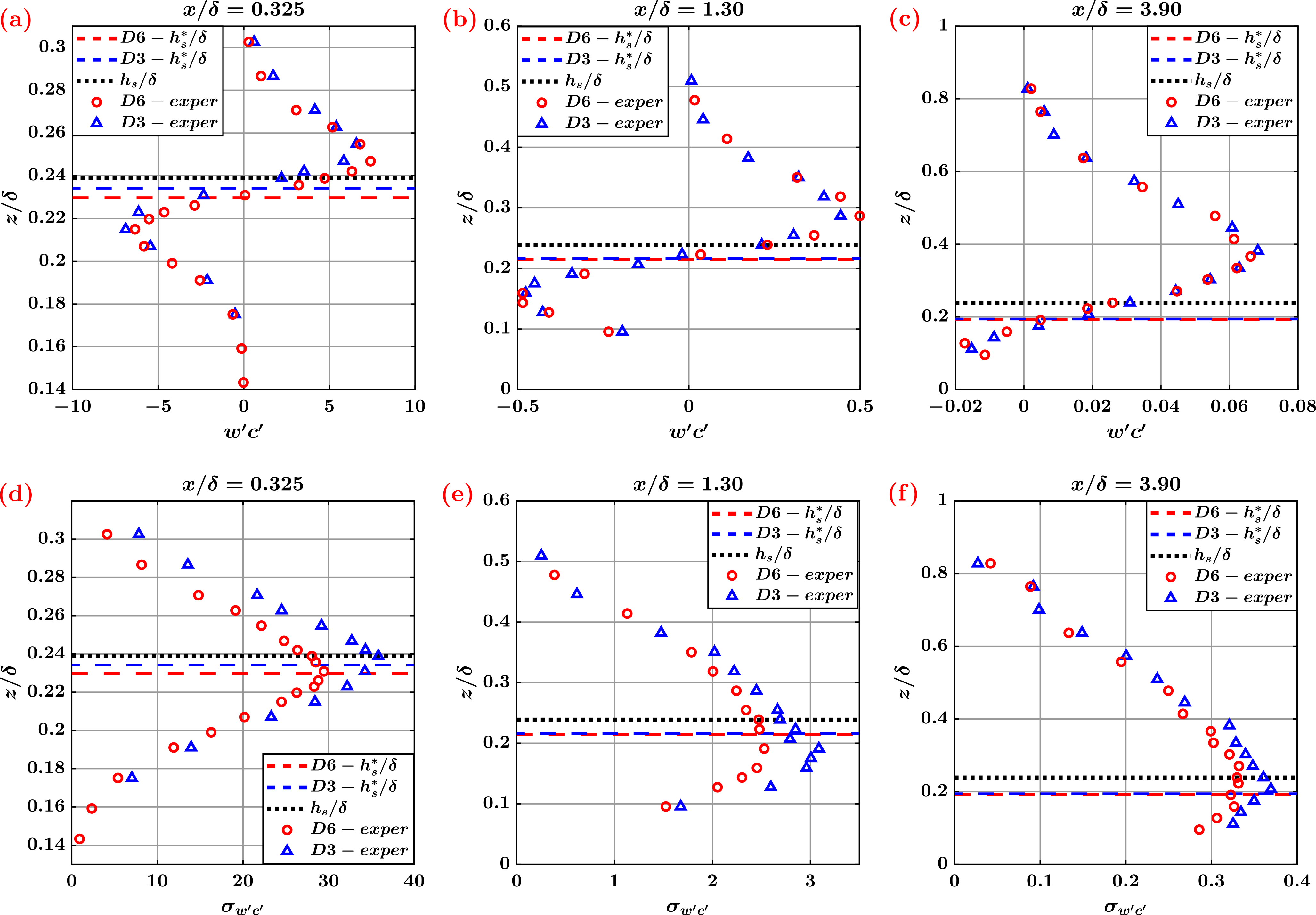}
			\caption{Vertical profiles of the mean value of wall-normal transport $\overline{w'c'}$ (a-c), and the standard deviation $\sigma_{w'c'}$ (d-f). The profiles are plotted at three streamwise locations, in the near field ($x/\delta=0.325$), in the far field ($x/\delta=3.90$), and at an intermediate location ($x/\delta=1.30$), for the two source diameters, $D=3$ mm and $D=6$ mm. The wall-normal coordinate of the source axis, $h_s$, is illustrated as a horizontal dotted line, while the plume axis height, $h_s^*$, is displayed as a blue (red) dashed line for the source D3 (D6).}
			\label{fig:mean_std_wc}		
		\end{figure*}

\section{Statistical analysis of the plume dynamics}	\label{sec:statistics}

		Previous works (e.g., \cite{fackrell1982concentration,nironi2015dispersion}) focused on the influence of the source size on the one-point concentration statistics, at varying distance from the source. Namely, it was shown that, while the mean concentration profiles are almost unaffected by the source conditions, the higher-order statistics (variance, skewness, kurtosis) show a high sensitivity on the source size, even at large distances from the release point \cite{nironi2015dispersion}. In a similar way, we examine here the statistics of vertical turbulent transport, $w'c'$. We show the vertical profiles of the statistics by focusing on the effect of the source size in the two configurations D3 and D6, corresponding to source diameters $D/\delta\approx9.55\times10^{-3}$ and $D/\delta\approx1.91\times10^{-2}$, respectively.
	
	\begin{figure*}[t]
				\centering
				\includegraphics[width=\linewidth]{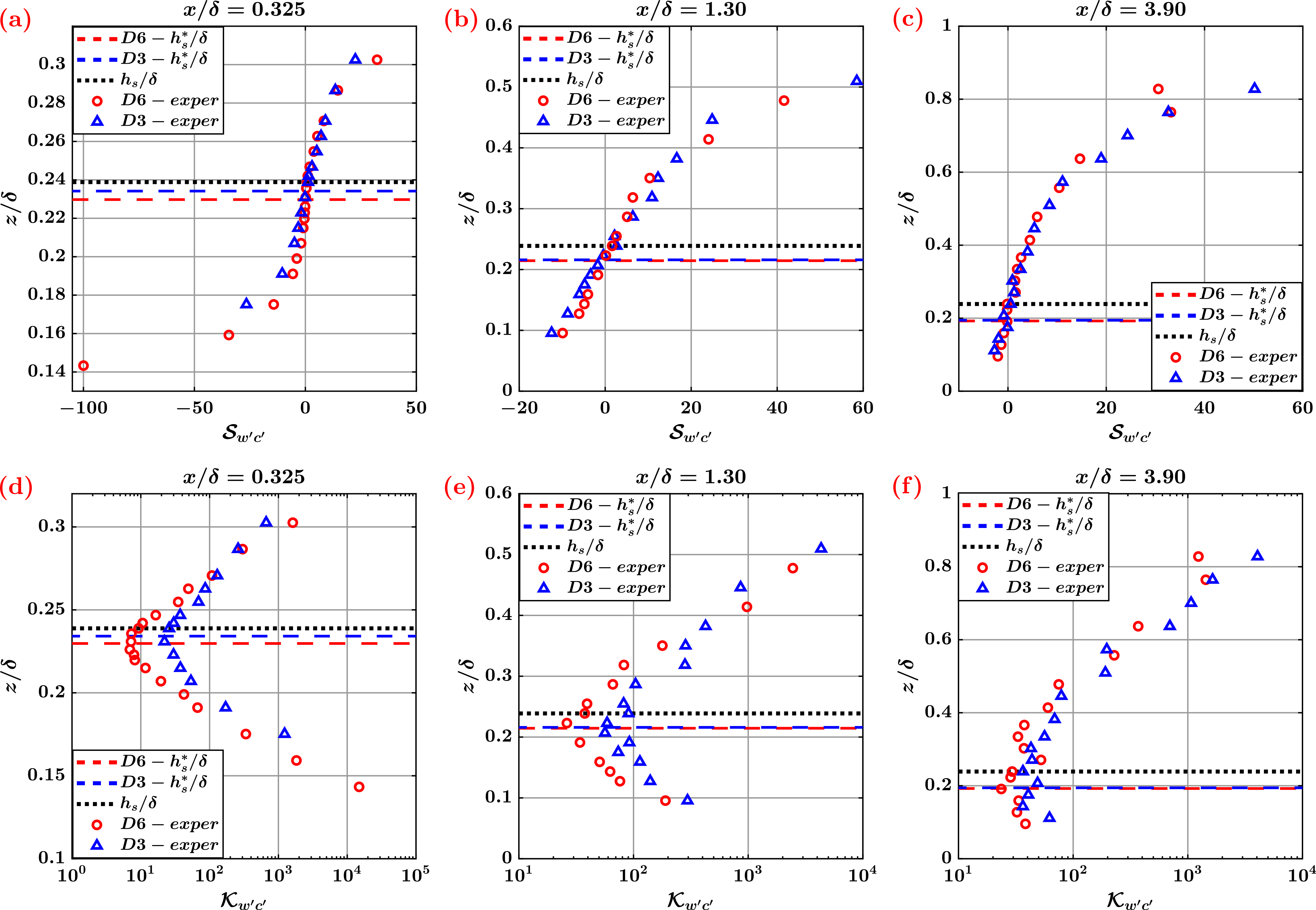}
				\caption{Vertical profiles of the skewness of wall-normal transport $\mathcal{S}_{w'c'}$ (a-c), in linear-linear scale, and the Kurtosis $\mathcal{K}_{w'c'}$ (d-f), in log-linear scale. The profiles are plotted at three streamwise locations, in the near field ($x/\delta=0.325$), in the far field ($x/\delta=3.90$), and at an intermediate location ($x/\delta=1.30$), for the two source diameters, $D=3$ mm and $D=6$ mm. The wall-normal coordinate of the source axis, $h_s$, is illustrated as a horizontal dotted line, while the plume axis height, $h_s^*$, is displayed as a blue (red) dashed line for the source D3 (D6).}
				\label{fig:sk_ku}		
			\end{figure*}		
				
	\subsection{Mean and standard deviation of turbulent transport}\label{subsec:mu_sigma}

			Fig. \ref{fig:mean_std_wc} shows the vertical profiles of mean and standard deviation values of the (normalized) turbulent flux, $w'c'$. The profiles are reported at three representative streamwise locations, i.e. in the near field ($x/\delta=0.325$), in the far field ($x/\delta=3.90$), and at an intermediate location ($x/\delta=1.3$). The vertical profiles of $\overline{w'c'}$ -- namely the total mass transport -- tend to collapse for the two configurations D3 and D6, as shown in Fig. \ref{fig:mean_std_wc}(a-c). This behaviour is more evident in the far field than in the proximity of the source, as the dependence of the mean concentration on the source size $D$ rapidly vanishes downstream from the source. In fact, the plume sizes -- i.e. the transversal and wall-normal spread of the plume --  become much more larger than $D$ with increasing $x/\delta$ due to the relative dispersion (e.g., see Fig. \ref{fig:metod}); as a result, the effect of $D$ on $\overline{w'c'}$ becomes negligible for $x/\delta\gg0$. The vertical turbulent transport is zero at the plume axis, $h_s^*$ (as an effect of the mean shear), namely where the vertical concentration gradient is minimum (see Fig. \ref{figApp:mean_c} in Appendix \ref{app:PAE}). Above and below the plume axis, instead, the mean value of turbulent flux is non-zero (see Fig. \ref{fig:mean_std_wc}(a-c)): for $z>h_s^*$, $\overline{w'c'}$ is positive while, for $z<h_s^*$, $\overline{w'c'}$ is negative. Above the source axis, the passive scalar is mainly carried out upwards by positive $w'$ fluctuations; on the other hand, below the source axis the passive scalar is mainly transported downwards by negative $w'$ fluctuations. In particular, the maximum/minimum value of $\overline{w'c'}$ corresponds to the maximum of the mean concentration gradient (this is also evident by using the Boussinesq approximation $\overline{w'c'}\sim-\partial\overline{c}/\partial z$ \cite{arya1999air}).
			
			As shown in Fig. \ref{fig:mean_std_wc}(d-f), the effect of the source size for an elevated source is instead much more evident for the standard deviation, $\sigma_{w'c'}$, rather than for the mean values -- in analogy with what is observed in the concentration statistics \cite{fackrell1982concentration,nironi2015dispersion} -- even at large distances from the source. This is a consequence of the stronger meandering motion of the plume emitted by the smallest source size, D3, which produces more variability in the series and the corresponding high intermittency in the dynamics of vertical turbulent transport (see Section \ref{subsec:spec_gamma}). The maximum difference of standard deviation between D3 and D6 is present close to the plume axis in the near field, and such difference strongly decreases by moving downstream towards the far field due to the weakening of the meandering and the strengthening of the relative dispersion. By moving in the wall normal direction, $\sigma_{w'c'}$ decreases as the plume intensity vanishes away from the source axis.

		\subsection{Skewness and kurtosis of turbulent transport}\label{subsec:sk_ku}

			The behaviour of the higher order moments is here investigated by focusing on the skewness, $\mathcal{S}_{w'c'}$, and the kurtosis, $\mathcal{K}_{w'c'}$, of the vertical turbulent flux. Formally, they are defined as the normalized third- and fourth-order central moments, namely $\mathcal{S}_{w'c'}=\overline{(w'c'-\overline{w'c'})^3}/\sigma_{w'c'}^3$ and $\mathcal{K}_{w'c'}=\overline{(w'c'-\overline{w'c'})^4}/\sigma_{w'c'}^4$, respectively. Fig. \ref{fig:sk_ku} shows the skewness and the kurtosis as a function of $z/\delta$ for three streamwise locations (as for the mean and standard deviation shown in Fig. \ref{fig:mean_std_wc}). 
			\\The behaviour of the skewness (Fig. \ref{fig:sk_ku}(a-c)) is similar for the two source configurations D3 and D6 at any $x/\delta$. In particular, $\mathcal{S}_{w'c'}\approx 0$ at the plume axis, while the skewness is negative/positive below/above the plume axis, because below and above $h_s^*$ the vertical turbulent transport is mainly downwards (i.e., $w'c'<0$) and upwards (i.e., $w'c'>0$), respectively. As shown in Fig. \ref{fig:sk_ku}(d-f), the kurtosis values are greater than three (which corresponds to normal distribution), thus implying that the PDFs of $w'c'$ are fat-tailed distributions (under some circumstances, the PDFs can well fitted by a Gamma distribution \cite{nironi2015dispersion}). In particular, $\mathcal{K}_{w'c'}$ is minimum at the plume axis at each stremwise location, as the plume develops around $z=h_s^*$ and extreme events (with respect to $\overline{w'c'}$) are less probable to appear; on the contrary, away from the plume axis extreme $w'c'$ values are more probable, as the signals are much more intermittent. Differently from the skewness, the kurtosis profiles for D3 and D6 are different in the near field (see Fig. \ref{fig:sk_ku}(d)), and the difference of $\mathcal{K}_{w'c'}$ progressively reduces towards the far field (see Fig. \ref{fig:sk_ku}(f)). This implies that the meandering affects the behaviour of the standard deviation but also the behaviour of the kurtosis: the values of $\mathcal{K}_{w'c'}$ for D3, in fact, are higher than the values of $\mathcal{K}_{w'c'}$ for D6, namely extreme values are more probable for D3 than for D6.

		\subsection{Spectra and intermittency factor} \label{subsec:spec_gamma}
		
			\begin{figure*}[t]
				\centering
				\includegraphics[width=.95\linewidth]{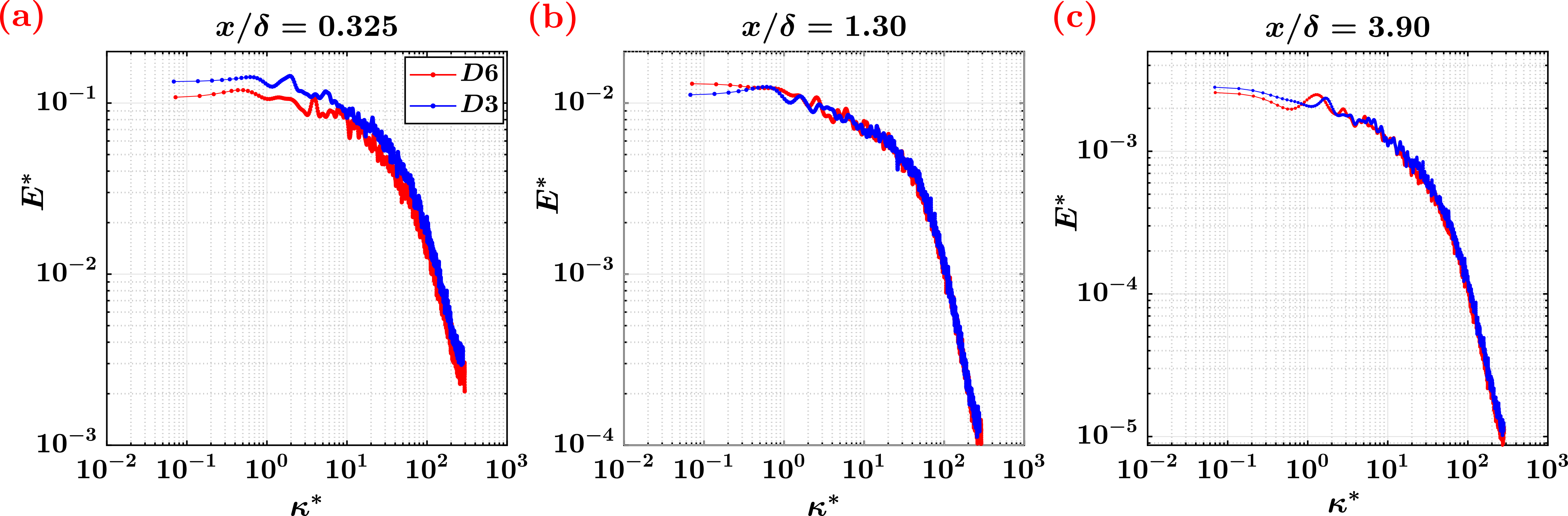}
				\caption{Normalized spectral density, $E^*$, as a function of the normalized wavenumber, $\kappa^*$, of the wall-normal turbulent flux, $w'c'$. Spectra are evaluated at the source axis ($z=h_s$), for the two source diameters, $D=3$ mm and $D=6$ mm, at three streamwise locations: (a) in the near field ($x/\delta=0.325$), (b) at an intermediate location ($x/\delta=1.30$), and (c) in the far field ($x/\delta=3.90$).}\label{fig:spettri}	
			\end{figure*}

			Fig. \ref{fig:spettri} shows the normalized power spectral density, $E^*=E\delta/\sigma_{w'c'}$, of the signals $w'c'$, as a function of the normalized wavenumber $\kappa^*=\kappa\delta$, where $\kappa=2\pi/\lambda$ is the wavenumber, $\lambda=\overline{u}/f$ is a characteristic turbulent length scale (see Fig. \ref{fig:metod}), $f$ is the frequency and $\overline{u}$ is the (local) mean streamwise velocity. Spectra are plotted along the source axis, namely for $z=h_s$. Since the (instantaneous) plume size, $\lambda_z$, depends on the source size and the spatial location, the relation between $\lambda_z$ and turbulent length scales, $\lambda$, affects the behaviour of the spectra. In particular, $\lambda_z$ is smaller for D3 than for D6 in the near field, due to the spatial proximity of the source; by moving downstream, instead, $\lambda_z$ increases and the difference of $\lambda_z$ between D3 and D6 diminishes. In the near field (Fig. \ref{fig:spettri}(a)), the difference of spectral density between D3 and D6 is larger at small wavenumbers than at high wavenumbers. In fact, turbulent length scales, $\lambda$, larger than the (instantaneous) plume size, $\lambda_z$, contribute to the (instantaneous) plume meandering motion in the wall-normal direction (e.g., see the sketch in Fig. \ref{fig:metod}). Therefore, since in the near field $\lambda_{z,\textrm{D3}}<\lambda_{z,\textrm{D6}}$, the differences of $E^*$ between D3 and D6 at low $\kappa^*$ are more evident, because the plume for D3 is affected by a wider range ($\lambda_z<\lambda<\lambda_{max}$, or equivalently, $\kappa^*_{min}<\kappa^*<2\pi\delta/\lambda_z$) of turbulent scales. On the other hand, at high wavenumbers (namely, small turbulent length scales), the spectral density for the two source sizes tends to coincide, as turbulent scales $\lambda<\lambda_z$ only promote the dispersion of the plume. The large scale fluctuations -- induced by a wider range of turbulent scale in the near field -- progressively weaken towards the far field (see Fig. \ref{fig:spettri}(b)-(c)), so that the intensity of spectral density decreases with $x/\delta$ and approaches the same behaviour at all wavenumbers for D3 and D6. In fact, for increasing $x/\delta$, the plume size increases (i.e., $\lambda_z\rightarrow\lambda_{max}$) and the range of scales for which $\lambda>\lambda_z$ decreases. In other words, all turbulent scales tend to contribute to the relative dispersion of the plume in the far field. Finally, it should be noted that spectra of vertical turbulent transport normalized by its variance do not show a self-similar behaviour along the streamwise direction. This is in contrast to what has been recently reported for concentration series, which show a self-similar behaviour in the range $0.5\leq x/\delta\leq 4$ \cite{talluru2019self}.		
		
			\begin{figure}[t]
				\centering
				\includegraphics[width=.95\linewidth]{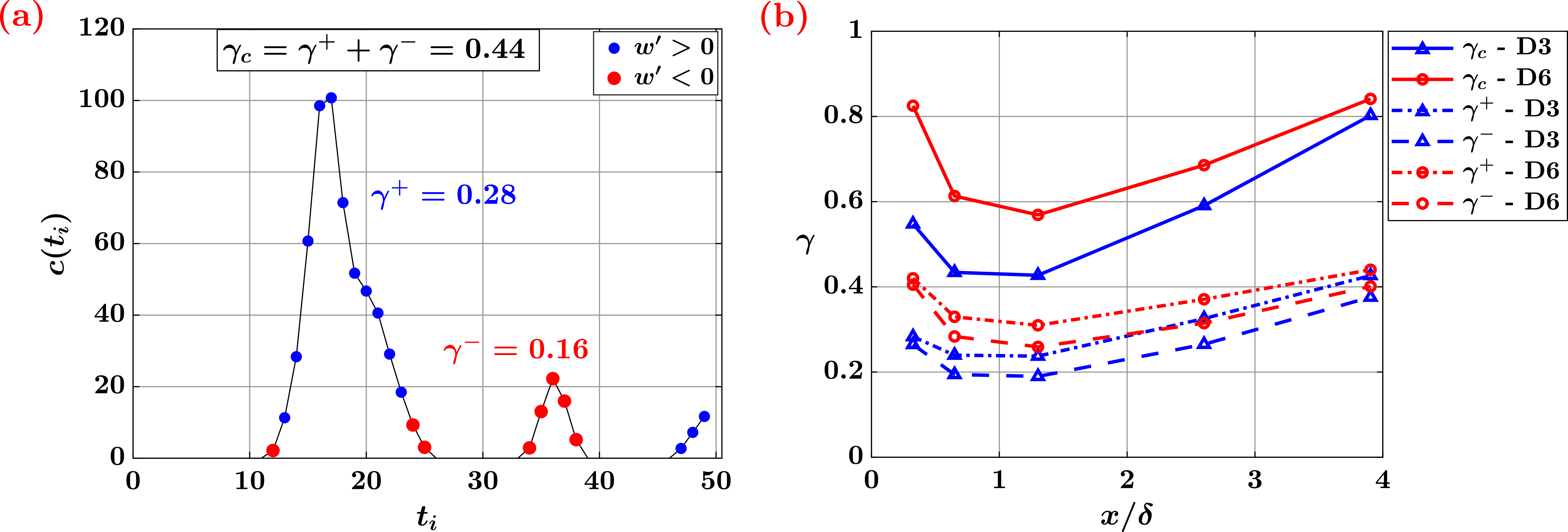}			
				\caption{(a) Example of the intermittent behaviour of the concentration signal (first 50 values) measured at $x/\delta=1.30$ and at the source axis. Blue and red data correspond to an upward and downward transport, respectively. The corresponding values of the intermittency factor for the shown time interval are also reported. (b) Intermittency factors of the passive scalar concentration, $\gamma_c$, and wall-normal turbulent flux, $\gamma^+$ and $\gamma^-$, as a function of $x/\delta$ along the source axis (i.e., $y/\delta=0$ and $z=h_s$).}
				\label{fig:intermitt}		
			\end{figure}
		
			The last parameter investigated is the intermittency factor. For the concentration series, an intermittency factor, $\gamma_c$, can be defined as the fraction of non-zero concentration values, where small $\gamma_c$ values correspond to highly intermittent series \cite{nironi2015dispersion}. In other words, $\gamma_c$ is the fraction of time in which the passive scalar is measured. In a similar way, here we define the intermittency factor for the vertical turbulent transport as the fraction of time in which the passive scalar is transported upwards, $\gamma^+=\textrm{prob}\left[w'>0,c\neq0\right]$, and downwards, $\gamma^-=\textrm{prob}\left[w'<0,c\neq0\right]$, with $\left(\gamma^++\gamma^-\right)=\gamma_c$ (by definition) and $\textrm{prob}\left[\bullet,\bullet\right]$ indicating the joint probability. In the definition of $\gamma^+$ and $\gamma^-$, the velocity fluctuations, $w'$, impose the sign to the fluxes while the concentration discriminates between the presence ($c\neq 0$) or the absence ($c=0$) of the plume. Therefore, although the overall intermittency is governed by the concentration field (i.e., $c\neq0$ or $c=0$), the velocity component introduces a directionality for the intermittency (namely $w'>0$ or $w'<0$). For example, the intermittency factor for the portion of the concentration signal shown in Fig. \ref{fig:intermitt}(a) is $\gamma_c=0.44$, because there are $22$ non-zero values of $c(t_i)$ out of $50$ values. Among the $22$ non-zero values, $14$ observations correspond to an upward motion (i.e., $\gamma^+=14/50=0.28$), while $8$ observations correspond to a downward motion (i.e., $\gamma^+=8/50=0.16$).
			
			Fig. \ref{fig:intermitt}(b) shows that, at the source axis, the vertical transport is more intermittent downward ($w'<0$) than upward ($w'>0$), namely $\gamma^-<\gamma^+$ for both D3 and D6. Consistently with the results shown in Fig. \ref{fig:mean_std_wc} and Fig. \ref{fig:sk_ku}, at the source axis the passive scalar is mainly transported upwards (as the plume axis lies below the source axis). Consequently, the fraction of time in which the passive scalar is transported upwards, $\gamma^+$, results to be greater than the fraction of time in which the passive scalar is transported downwards, $\gamma^-$. Furthermore, the intermittency factor is always smaller for the source diameter D3 than for D6, whether $\gamma_c$, $\gamma^+$ or $\gamma^-$ is considered. This validates the fact that meandering motion is stronger for the plume emitted by a smaller source, inducing higher intermittency in the signals. More in detail, in the near field, the values of intermittency factor for D3 and D6 are different while they approach the same value in the far field, as the effect of the meandering is replaced by the relative dispersion of the plume. Although a strong meandering motion is present in the near field, the intermittency factors do not monotonically increase with $x/\delta$ because of the effect of the source proximity in the near field (as mentioned in Section \ref{subsec:mu_sigma} for the mean turbulent transport). Therefore, a minimum value of the intermittency is found at $x/\delta\approx1.3$, which is also found by \textit{Nironi et al.} \cite{nironi2015dispersion} in the case of $\gamma_c$.

\section{Visibility-network analysis of turbulent transport} \label{sec:visib}
	
	In this section, we present the results of the analysis of the vertical turbulent transport by means of the visibility graph approach. The network analysis is here performed to advance the level of information of classical statistical analysis (see Section \ref{sec:statistics}), thus providing a richer picture of the plume dynamics via turbulent transport investigation. Differently from classical statistics tools, different temporal arrangements of the same time-series generate different visibility networks (e.g., a shuffled series maintains the same statistics of the originating series, but exhibits a different temporal structure and visibility-network). Since the network metrics evaluated from the visibility graph approach are able to characterize the temporal structure of the time-series, they carry high-order and nonlinear information of the signal \cite{iacobello2018visibility}. As a result, the visibility graph approach is proposed to shed light on the temporal structure -- in terms of extreme events and their relative intensity -- of the turbulent transport time-series.
	
	\subsection{Concepts and definitions}\label{subsec:net_def}
	
		A complex network is formally defined as a graph that shows non-trivial topological features \cite{newman2018networks}. Networks are made up of $N$ entities called \textit{nodes} interconnected by a set of $L$ \textit{links}, and they are commonly represented by an \textit{adjacency matrix}, which is defined as 
	\begin{equation}\label{eq:Aij}
		A_{ij} = \left \{\begin{array}{l} 1$, if $\lbrace i,j\rbrace\in\mathcal{L}$, with $i\neq j,\\ 0$, otherwise$, 
			\end{array}	
			\right.
	\end{equation}
	
	\noindent where	$i,j=1,...,N$ and $\mathcal{L}$ is the set of $L$ links. Therefore, the entries $A_{ij}$ take into account the presence of a link between each pair of nodes. In this work, we considered each connection to be undirected (i.e., $A_{ij}=A_{ji}$) and unweighted, resulting into binary and symmetrical adjacency matrices. 
			
\begin{figure*}[t]
		\centering
		\includegraphics[width=\linewidth]{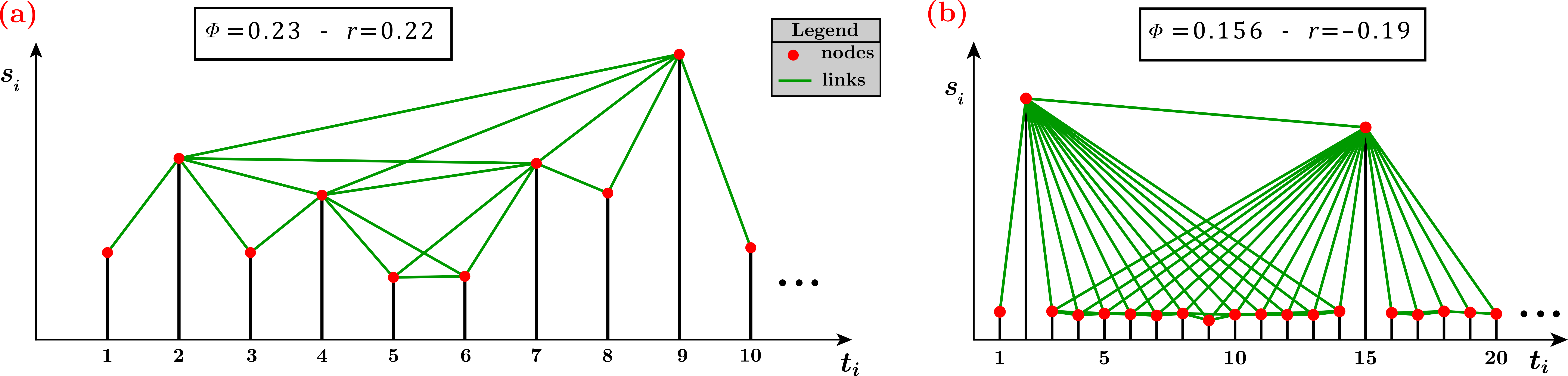}
		\caption{Examples of two intervals of time-series, $\left(t_i,s_i\right)$, and corresponding visibility networks, showing different temporal structures. Nodes and links are depicted as red dots and green lines, respectively. (a) First 10 out of $10^5$ observations of a series extracted from a uniform probability distribution in the interval $\left[0,1\right]$. (b) First 20 out of $10^5$ observations of a series extracted from a uniform probability distribution in the interval $\left[0,1\right]$, with periodic spikes (every 14 instants, $t_i$) uniformly distributed in the interval $\left[0,100\right]$. The values of $s_i$ in the vertical axis are not shown due to the invariance of visibility algorithm to affine transformations.}
		\label{fig:eg_vis}		
	\end{figure*}

		In order to investigate the time-series of turbulent transport, we employed the natural visibility graph (NVG) method \cite{Lacasa2008}, which is a widely exploited technique to map time-series in complex networks. This method was firstly proposed by \textit{Lacasa et al.} \cite{Lacasa2008}, who showed that the resulting visibility-network is able to inherit important features of the mapped time-series. According to the algorithm, each datum of a time-series, $s_i\equiv s(t_i)$, is mapped in a node of the network and a link between two nodes, $(i,j)$, is established if:
	
	\begin{equation}\label{eq:visib}
		s_k<s_j+\left(s_i-s_j\right)\frac{t_j-t_k}{t_j-t_i},
	\end{equation}
	
	\noindent for all $t_k$ between $t_i$ and $t_j$ (or analogously $\forall k, i<k<j$). Therefore, by construction, each node $i$ is always linked to its immediately closest nodes, namely $j=i\pm 1$. From Eq. (\ref{eq:visib}) it follows that the natural visibility algorithm satisfies a convexity criterion, so that (subsets of) nodes that form a convex series (e.g., a bowl-like series) are fully linked with each other. In this work, each network has $N=N_T=1.8\times10^5$ nodes, corresponding to the recorded data values of each velocity and concentration series. The most important nodes (\textit{hubs}) for a visibility-network are associated with positive peaks in the series, because very high values are more likely to see other nodes (i.e., hubs have a better visibility). Fig. \ref{fig:eg_vis} shows two simple examples of time-series (illustrated as black stems) mapped into visibility networks, where nodes and links are depicted as red dots and green lines, respectively. It is worth highlighting that the visibility criterion emphasizes the positive peaks, but not the negative ones. Consequently, when the series mainly display \textit{pits} (i.e., negative peaks) instead of positive peaks, it is possible to exploit the Eq. (\ref{eq:visib}) to build visibility networks from the complementary series, $-s_i$. The comparison of the metrics extracted from the original series, $s_i$ and its opposite, $-s_i$, allows one to characterize the peak-pit asymmetry in the series \cite{hasson2018combinatorial}, namely if peaks are mainly positive or negative.
	
		The main advantage of the visibility algorithm relies on the fact that it does not require any \textit{a priori} parameter. However, the NVG is invariant under rescaling and translation of both horizontal and vertical axes (i.e., affine transformations) \cite{Lacasa2008}. This peculiar feature of NVG is crucial, and it must always be taken into account when the network structure is investigated. In fact, the invariance to affine transformations implies that two time-series with different mean and standard deviation values but with similar temporal structure, are mapped into the same visibility network. If the analysis should be sensitive to affine transformations of the series, the invariance of NVG represents a drawback of the method. On the other hand, if the focus is on the temporal structure of the series (as in this work), the invariance is a potential benefit. In fact, it is possible to exploit the NVG to analyse series without a pre-processed normalization of the series. Finally, it should be noted that the condition in Eq. (\ref{eq:visib}) still holds for non-uniform sampling or time-series with missing values \cite{zou2018complex}, which is an advantage when incomplete data measurements are available.

	\subsection{Network metrics}\label{subsec:net_metr}
	
In order to characterize the structure of complex networks, several metrics have been proposed so far \cite{newman2018networks}. In a previous work on the investigation of velocity time-series in a turbulent channel flow, three metrics were exploited \cite{iacobello2018visibility}: the transitivity, the mean link-length and the degree centrality. Among these three metrics, the transitivity and the mean link-length are able to highlight the presence of small variations in the series and the occurrence of peaks, respectively. On the other hand, the degree centrality cannot be uniquely related to a specific temporal feature, since the degree accounts for both the recurrence of peaks and the presence of small variations in the series \cite{iacobello2018visibility}. Therefore, to extract information on the characterization of extreme events of the plume dynamics we here exploited two specific network metrics: the \textit{mean link-length}, and the \textit{assortativity coefficient}. The mean link-length, $d_i$, was proposed by \textit{Iacobello et al.} \cite{iacobello2018visibility} as a local measure to characterize the temporal occurrence of peaks in a series, defined as
	
		\begin{equation} \label{eq:d1n_i}
			d_i=\frac{1}{k_i} \sum_{j=1}^N{|t_j-t_i|A_{ij}},	
		\end{equation}
		
	\noindent where $k_i=\sum_j{A_{ij}}$ is the degree of a node $i$, that is the number of nodes linked to $i$. The average value over all nodes of the mean link-length is then $\langle d\rangle=\sum_i{d_i}/N$, and it represents a characteristic temporal distance between two visible data in a series. A high value of the average mean link-length, $\langle d\rangle$, indicates that peaks rarely occur in a series, since peaks prevent visibility between data that are far from each other \cite{iacobello2018visibility}. Therefore, large $d_i$ values correspond to hubs in the network and peaks in the series, as peaks activate long-range links (i.e. farther temporal \textit{horizons}). As mentioned above, although the degree is usually adopted as the metric to characterize hubs, the mean link-length reveals to be a more reliable metric than degree for peaks characterization in visibility networks. For example, in a visibility-network built on a fully convex series (e.g., a bowl-like series with a minimum value), each node has maximum degree value equal to $N-1$ due to the full convexity of the series, but the node corresponding to the minimum value does not represent a peak. It is also worth noting that while the kurtosis is an estimation index of extreme values in a PDF, the mean link-length quantifies the average temporal distance between extreme events (while the PDF is invariant to the temporal structure of the signal). In this work, in order to better highlight the cases in which peaks frequently appear (i.e., low $\langle d\rangle$ values), we introduce the \textit{average peak occurrence}, $\phi=\langle d\rangle^{-1}$, corresponding to a characteristic frequency of peaks in a series. 	
	
	It must be emphasized that we refer to peaks as the local (or global) highest positive values in the series \cite{iacobello2018visibility}. This does not necessarily imply that peaks also correspond to outliers, namely very large values with respect to a local subset of data; on the contrary, outliers typically correspond to peaks. For example, in Fig. \ref{fig:eg_vis}(a) outliers are not present and peaks correspond to nodes $i=\{2,4,7,9\}$; in Fig. \ref{fig:eg_vis}(b), instead, peaks correspond to outliers, namely nodes $i=\{2,15\}$. Therefore, the average peak occurrence, $\phi$, is sensitive to the appearance of peaks in time (horizontal separation), but it does slightly take into account the relative intensity of peaks compared to all the other values in the series. To address this issue, we also investigated the assortativity coefficient, $r$, which is the Pearson correlation coefficient of the degree of the nodes at the ends of each link \cite{newman2002assortative}, namely

	\begin{equation} \label{eq:assort}
		r=\frac{\sum_i{\sum_{j>i}{\frac{A_{i,j}}{L}k_i k_j}}-\left[\sum_i{\sum_{j>i}{\frac{A_{i,j}}{2L}(k_i+k_j)}}\right]^2}{\sum_i{\sum_{j>i}{\frac{A_{i,j}}{2L}(k_i^2+k_j^2)}}-\left[\sum_i{\sum_{j>i}{\frac{A_{i,j}}{2L}(k_i+k_j)}}\right]^2},
	\end{equation}
	 \noindent where $(i,j)$ are the end-nodes of each link $l\in\mathcal{L}$. Positive $r$ values are obtained when nodes are linked with other nodes of similar degree: in this case, the network is said \textit{assortative}. On the contrary, the network is said \textit{disassortative} if $r<0$, or \textit{non-assortative} if $r=0$. As for the correlation coefficient, $r$ ranges in the interval $\left[-1,1\right]$. In particular, a negative $r$ value means that high degree nodes (i.e., nodes with more visibility) tend to be more linked with low degree nodes (i.e., nodes with less visibility), rather than with other high degree nodes. When peaks are focused, the assortativity coefficient quantifies the extent to which peaks (which are expected to have more visibility) are more prominent with respect to small fluctuations (which are expected to have less visibility). Highly positive values of $r$ indicate that peaks are slightly pronounced with respect to the other values in the series (e.g., Fig. \ref{fig:eg_vis}(a)), while strongly negative values of $r$ indicate a substantial presence of outliers (e.g., Fig. \ref{fig:eg_vis}(b)). In other words, $r$ is a measure of the vertical separations in the series, that is how intense peaks are with respect to the other data in the time-series. As a rule of thumb, $r=0$ discriminates between the prominence of peaks (i.e, $r> 0$) and the prominence of outliers (i.e., $r< 0$). If the network is non-assortative (i.e., $r\approx 0$), in general neither peaks nor outliers are expected to be prominent. However, if two signals are compared, it can be inferred that the signal showing $r\approx 0$ is more likely to have outliers or peaks than the signal with $r>0$ or $r<0$, respectively.
	 
	 To summarize, the two network metrics, $\phi$ and $r$, are here selected to characterize the temporal structure of the series in terms of horizontal and vertical separations, respectively. For instance, in Fig. \ref{fig:eg_vis}(a), ${\phi=0.230}$ and $r=0.22$, while in Fig. \ref{fig:eg_vis}(b) ${\phi=0.156}$ and ${r=-0.19}$. These values are evaluated as described in the caption of Fig. \ref{fig:eg_vis}. Accordingly, $\phi$ is more reliable to detect the occurrence of peaks in the series, while $r$ is able to discern between peaks (in which $r$ is generally positive) and outliers (in which $r$ is generally negative) in a time-series. For example, large values of $\phi$ indicate that the corresponding series has many peaks, which appear as outliers only if $r$ decreases towards negative values. Accordingly, the two metrics, $\phi$ and $r$, should be analysed in pairs in order to infer the temporal structure of the series evaluated at different spatial locations.

		\begin{figure*}[t]
			\centering
			\includegraphics[width=\textwidth]{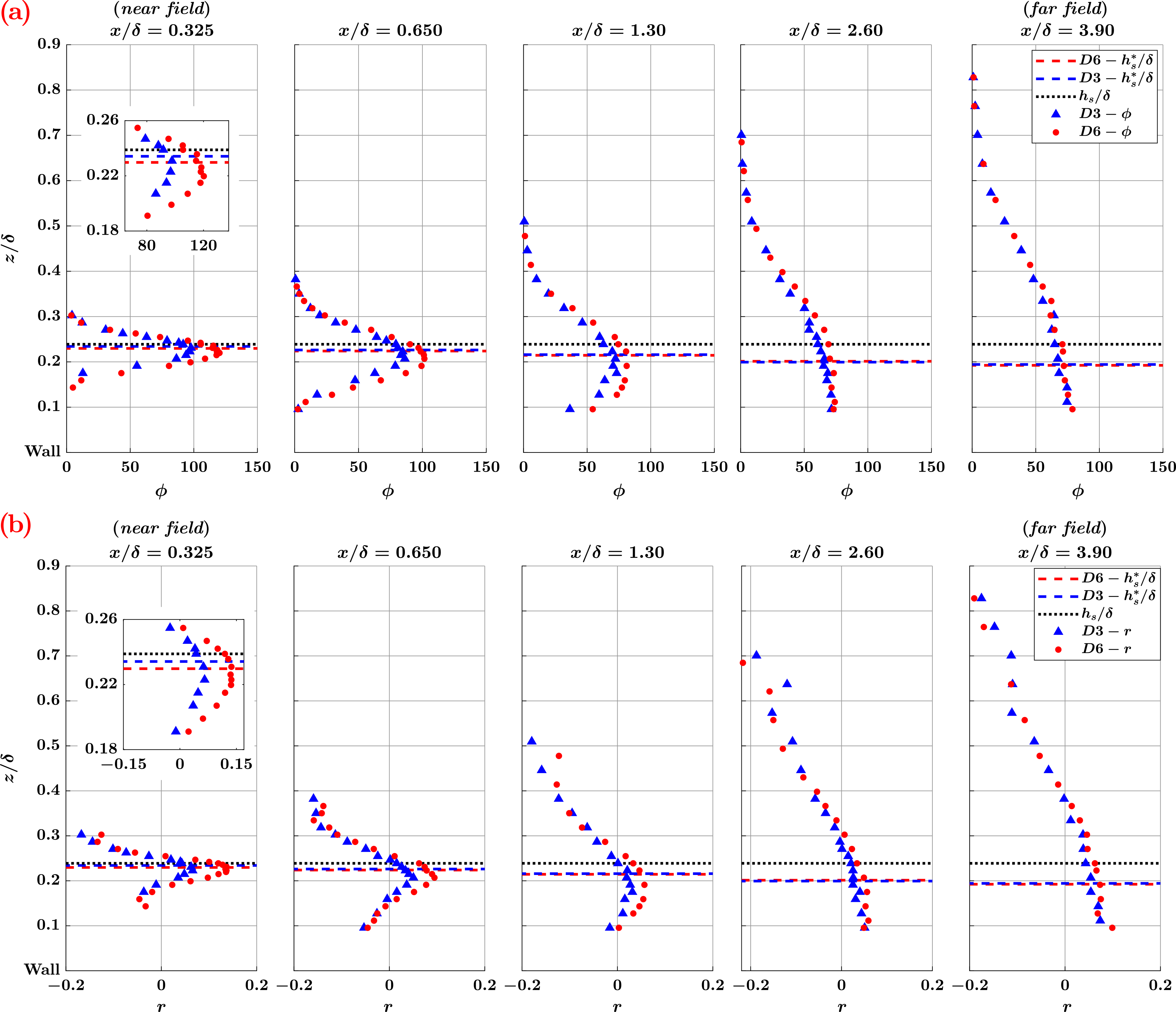}
			\caption{Vertical profiles of the average peak occurrence $\phi \left[ \textrm{Hz}\right]$ (a), and the assortativity coefficient $r$ (b), based on signals of wall-normal turbulent transport, $w'c'$. The profiles are plotted from the near field ($x/\delta=0.325$) to the far field ($x/\delta=3.90$), for the two source diameters, $D=3$ mm and $D=6$ mm. The wall-normal coordinate of the source axis, $h_s$, is illustrated as a horizontal dotted line, while the plume axis height, $h_s^*$, is displayed as a blue (red) dashed line for the source D3 (D6). The insets at $x/\delta=0.325$ show two zooms around the plume axis.}
			\label{fig:metriche_wc_top}
		\end{figure*}

\subsection{Spatio-temporal investigation of $w^{\prime}c^{\prime}$ through the network metrics}\label{subsec:metr_analysis}
	
		The results from the application of the visibility algorithm to the time-series of turbulent transport, $w'c'$, are reported in this Section. The two configurations of source diameter, D3 and D6, are displayed for different downstream locations, $x/\delta$, and at various wall-normal coordinates, $z/\delta$. 
		
\begin{figure*}[t]
			\centering
			\includegraphics[width=\linewidth]{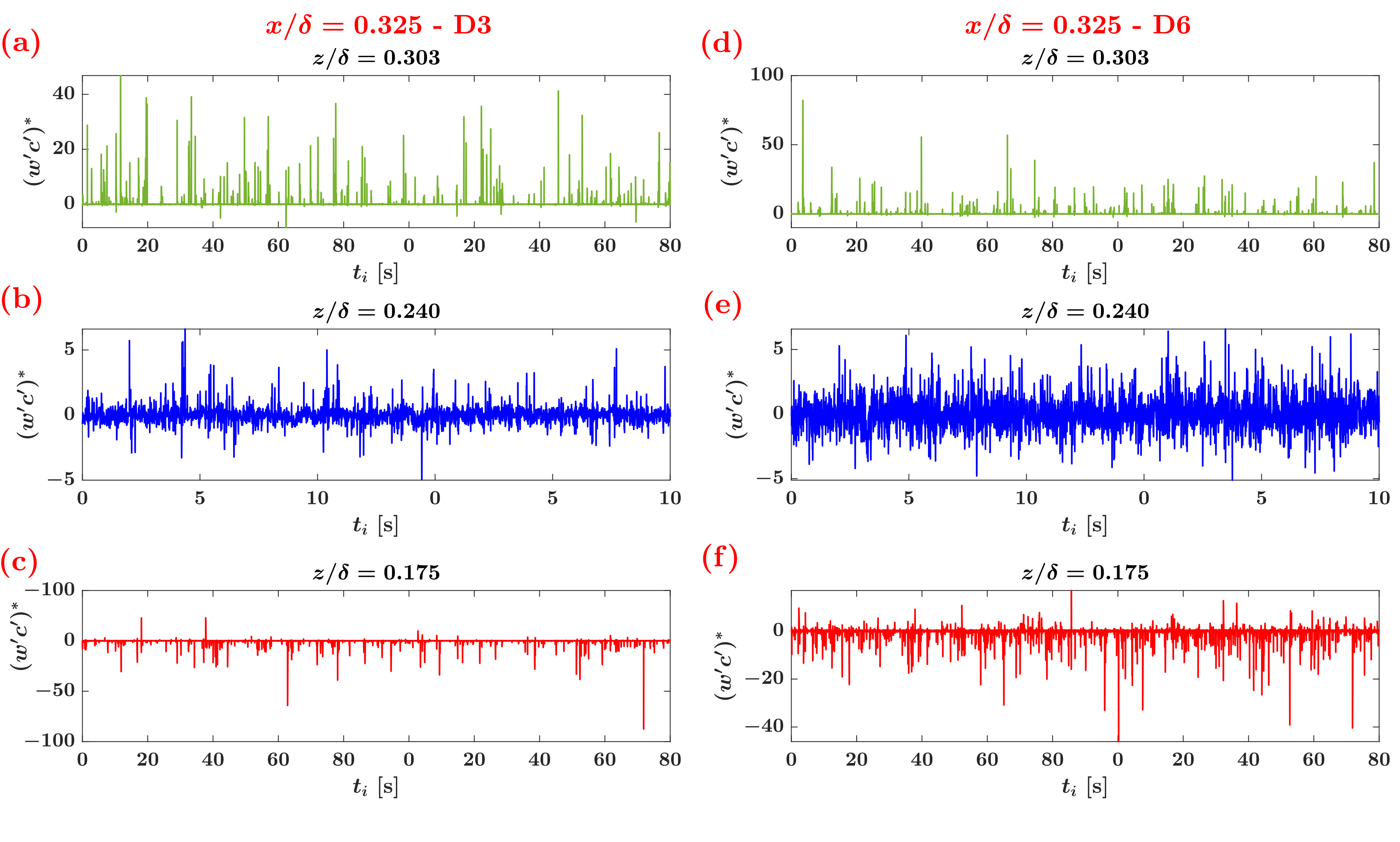}
			\caption{Time-series of vertical turbulent transport in the near field, $x/\delta=0.325$, for the source diameter D3 (a-c) and D6 (d-f). Signals are plotted at three wall-normal coordinates, namely above the source axis at $z/\delta=0.303$ (panels (a) and (d)), at the source axis $h_s/\delta=0.240$ (panels (b) and (e)), and below the source axis at $z/\delta=0.175$ (panels (c) and (f)). For comparison purposes, the time-series are normalized as $(w'c')^*=\left(w'c'-\overline{w'c'}\right)/\sigma_{w'c'}$.}\label{fig:serie_wc_D3D6}
		\end{figure*}

		The behaviours of the average peak occurrence, $\phi$, and the assortativity coefficient, $r$, are shown in Fig. \ref{fig:metriche_wc_top}(a) and Fig. \ref{fig:metriche_wc_top}(b), respectively, as a function of the vertical coordinate, $z/\delta$. In general, $\phi$ and $r$ have their maximum values close to the plume axis, $h_s^*$ (see horizontal dashed lines in Fig. \ref{fig:metriche_wc_top}): this means that, around the plume axis, peaks frequently occur and the vertical separation between them and the other data in the series is weak (since $r>0$). This behaviour of the metrics is due to the fact that the plume is mainly located around the source axis while it meanders and develops downstream. The difference in the features of temporal structure of the $w'c'$ can be observed in Fig. \ref{fig:serie_wc_D3D6}, which shows segments of time-series of $w'c'$ measured at $x/\delta=0.325$ for D3 (Fig. \ref{fig:serie_wc_D3D6}(a-c)) for D6 (Fig. \ref{fig:serie_wc_D3D6}(d-f)). Specifically, high values in the signals are much more frequent (i.e., high $\phi$) and much less prominent (i.e., high $r$) around the plume axis (see Fig. \ref{fig:serie_wc_D3D6}(b,e)) than away from it.
		\\The effect of the source size, $D$, on the metrics is clearly visible in the near field (i.e., $x/\delta=0.325$) and also at intermediate streamwise distances (i.e., up to $x/\delta=1.30$) around the plume axis, where the metrics for D3 have smaller values than the metrics for D6. The peaks tend to appear more as outliers (i.e., lower $r$) and they occur less frequently (i.e., lower $\phi$) for the smallest source size D3 (see Fig. \ref{fig:serie_wc_D3D6}(b)) than for D6 (see Fig. \ref{fig:serie_wc_D3D6}(e)). Since a plume emitted from a smaller source diameter is affected by a wider range of turbulent scales, its meandering motion is more intense. A strong meandering motion implies a high variability (i.e., standard deviation) and a large fraction of small data values (i.e., strong intermittency), which are captured as different values between D3 and D6 of the network metrics, $\phi$ and $r$. It should be noted that the networks corresponding to locations around the plume axis do not show large negative $r$ values, as the higher variability in the signals prevent a strong separation between small values and peaks (which would give $r<0$). In the far field (i.e., $x/\delta=2.60-3.90$), the vertical profiles of the two metrics tend to collapse for both source diameter configurations, D3 and D6. This behaviour is a consequence of the increase of the plume size by moving downstream, due to the relative dispersion. In the far field the plume size exceeds the largest turbulent scales and the mixing of the passive scalar is then fully regulated by the relative dispersion rather than by the meandering. Therefore, by moving downstream around the plume axis, the average peak occurrence always decreases with $x/\delta$, since large concentration values (and in turn high turbulent transport) are less probable to appear in the far field because of the plume weakening. On the other hand, for $z\approx h_s^*$, the assortativity coefficient first decreases (reaching a minimum at about $x/\delta=1.30$) and then increases with $x/\delta$. The behaviour of the assortativity coefficient is still a consequence of the interplay between the reducing meandering motion and increasing dispersion of the plume as it evolves downstream. It is worth noting that the minimum value of $r$ along the source axis is found at $x/\delta\approx1.3$, that is the streamwise location where the plume reaches the ground and all the intermittency factors are minimum (see also Fig. \ref{fig:intermitt}(b)).

		By focusing on the effect of the wall normal coordinate, $z/\delta$, at a given streamwise location, the two metrics tend to decrease by moving away from the plume axis in the wall normal direction. Low values of $\phi$ and $r$ indicate that, from the point of view of the temporal structures of the series, peaks occur less frequently and such peaks appear as outliers (e.g., see Fig. \ref{fig:serie_wc_D3D6}(a,d) and Fig. \ref{fig:serie_wc_D3D6}(c,f) for series above and below the plume axis, respectively). Therefore, the temporal structure of the signals becomes spike-like moving away from the plume axis along the $z$ direction (see Fig. \ref{fig:serie_wc_D3D6}(a,c) for D3 and Fig. \ref{fig:serie_wc_D3D6}(d,f) for D6). Additionally, the metrics rapidly decrease with $z/\delta$ in the near field (as expected), since the plume rapidly vanishes in the vertical direction; in the far field, instead, the metrics gently decrease with $z/\delta$ from $z=h_s^*$, as the plume size increases downstream due to the relative dispersion. It is worth noting that, in the near field, the difference of average peak occurrence, $\phi$, between D3 and D6 is larger below than above the plume axis (e.g., see the first-left panel of Fig. \ref{fig:metriche_wc_top}(a)). Therefore, the behaviour of $\phi$ in the near field indicates that peaks in the series of D6 are more frequent than for D3 below the plume axis. This behaviour can be seen, for instance, by looking at the signals shown in Fig. \ref{fig:serie_wc_D3D6}(c,f). Since in the near field the effect of the source size is still notable, the plume for D6 is larger and more tilted towards the wall than for D3 (i.e., $h_s^*$ is smaller for D6). As a consequence, at a given vertical coordinate below the plume axis (e.g., at $z/\delta=0.175$ in Fig. \ref{fig:serie_wc_D3D6}) the temporal structure of the series for D6 (e.g., Fig. \ref{fig:serie_wc_D3D6}(f)) is less spike-like than for D3 (e.g., Fig. \ref{fig:serie_wc_D3D6}(c)). It should be emphasized that the behaviour of the network metrics along the wall-normal direction, ${z/\delta}$, is consistent with the outcomes reported in Ref. \cite{talluru2018local}, which provides a phenomenological description of the velocity–scalar interaction.

		Finally, we remark that the visibility algorithm emphasizes the presence of positive peaks in the signals. However, as shown in Fig. \ref{fig:serie_wc_D3D6}, there is an asymmetry between positive and negative extreme values in the series, especially away from the plume axis. Therefore, the analysis of $\phi$ and $r$ can also be carried out by applying the visibility algorithm to the signals $-w'c'$, thus highlighting the temporal occurrence and relative intensity of the negative extreme values (i.e., pits) of vertical turbulent transport. We found that the metrics obtained from $w'c'$ and $-w'c'$ coincide at the plume axis, $h_s^*$, while they are different away from it. In particular, above the plume axis the values of $\phi$ and $r$ are higher for the series $-w'c'$ than for $w'c'$, that is peaks are less frequent and appear more as outliers than pits; on the contrary, the opposite behaviour is found below the plume axis (see Appendix \ref{app:PPATTS} for more details).

	\section{Discussion and conclusions}\label{sec:conclus}

			The dispersion of a passive scalar in a turbulent boundary layer is experimentally investigated via wall-normal turbulent transport time-series, $w'c'$, for two elevated source sizes. The plume dynamics is analysed by means of classical statistical tools and through a complex network-based approach. The statistical analysis reveals that the mean value and the skewness of $w'c'$ are not substantially affected by the source size, while the standard deviation and the kurtosis are sensitive to the emission conditions. The plume meandering -- that is mainly active in the near field of the source -- is the main responsible for the differences in the statistics: a stronger meandering motion produces higher variability (i.e., standard deviation) and more extreme events (i.e., kurtosis) for the smallest source size than for the largest source size. Far from the source, the meandering motion strongly reduces its intensity and the relative dispersion turns out to be the principal mechanism affecting the plume dynamics. In the far field, therefore, the statistics approach the same values for the two source sizes. The effect of the meandering on turbulent transport is also highlighted by the power spectral density and the intermittency factor. Larger values of power spectral densities are obtained at small wavenumbers for the smallest source size, as a consequence of a wider range of turbulent scales affecting the plume dynamics. Additionally, the stronger meandering motion associated with the plume emitted by a smaller source size induces a more intense intermittency in the signals, namely a lower fraction of time for which the passive scalar is measured and transported. As for the statistical moments, the main differences of power spectral density and intermittency between the two source sizes are large in the source proximity and vanish in the far field. In this way, by investigating the vertical turbulent transport time-series, we extend the benchmark of (one-point) statistics of \textit{Nironi et al.} \cite{nironi2015dispersion} and \textit{Fackrell and Robins} \cite{fackrell1982concentration} on the dynamics of a passive scalar plume emitted in a rough-wall turbulent boundary layer.
		
		With the aim to advance the level of information of classical statistics, a complex network analysis is also carried out by exploiting the visibility algorithm. We focused on two network metrics: the average peak occurrence and the assortativity coefficient, which characterize the temporal structure of $w'c'$ in terms of occurrence and relative intensity of (positive) peaks, respectively. The network metrics are significantly affected by the measurement locations. Specifically, at each streamwise coordinate, the temporal structure of the signals at the plume axis is made up of peaks that frequently appear in time (i.e., high values of the average peak occurrence), while peaks appear less frequently away from the plume axis (i.e., small values of the average peak occurrence). The assortativity coefficient indicates that the relative intensity of peaks (with respect to all the other values in the signal) is smaller at the plume axis (positive assortativity) rather than away from it (negative assortativity): below and above the plume axis, therefore, peaks appear infrequently and as outliers. At the borders of the plume, indeed, the turbulent transport is reduced because of the presence of small concentrations (due to the relative dispersion of the plume), thus outliers sporadically appear in the corresponding time-series of turbulent transport.	
		\\By focusing on the effect of the source size, the metrics at the plume axis for the smallest source (D3) display lower values than the metrics for the largest source (D6). This means that the temporal structure of the $w'c'$ signals for D3 (i.e., for a stronger plume meandering) is characterized by positive peaks that appear more as outliers and are less frequent than for D6 (i.e., for a weaker plume meandering). Such difference between D3 and D6 is evident in the proximity of the source, while it decreases for increasing streamwise coordinates. Therefore, the unequal intensity of the meandering motion between D3 and D6 differently affects the temporal structure of turbulent transport series in the near field. Since a strong meandering motion induces a high intermittency in the series, it is expected that two peaks of turbulent transport are more unlikely to appear close in time (as there must be large fraction of time in which the passive scalar is not measured). Furthermore, the high variability (i.e., standard deviation) associated with the meandering motion implies that the relative intensity of peaks with respect to the other values in the signal increases (i.e., extreme events appear more as outliers), because the time-series between two peaks is made up of values that are different from the peaks (otherwise the standard deviation values would be small). 
 
		To summarize, the respective influence of the meandering motion and the relative dispersion for different source sizes, as well as the plume weakening with the streamwise distance from the source, is fully captured by the network metrics. Specifically, the average peak occurrence and the assortativity coefficient are able to highlight the different temporal structures of the series -- in terms of peaks occurrence and their relative intensity -- at different spatial locations. In this way, the visibility network-based approach enriches the analysis of the plume dynamics carried out via classical statistics, revealing significant information about the temporal structure of the measured signals. In fact, the network metrics are able to discern between peaks and outliers and their frequency in the signals, which cannot be easily deduced from classical statistics. As a result, the visibility algorithm can be exploited as an effective tool for the time-series analysis of experimental data, by highlighting non-trivial features of the main mechanisms affecting the plume dynamics.

\begin{acknowledgments}
	 P. Salizzoni acknowledges funding from the "R\'egion Auvergne-Rh\^one-Alpes" for the "SCUSI" project. The authors would also like to express their gratitude to Patrick M\'ejean and Horacio Correia for the technical support in the wind tunnel experiments.
\end{acknowledgments}

\appendix
\section{Characterization of the concentration and velocity field}\label{app:DET}
	
		Velocity time-series were acquired by means of a $\pm 45^\circ$ X-probe hot-wire anemometer (HWA) working at a constant temperature, which allows for the simultaneous measurements of two velocity components. The probe was calibrated by exploiting a Pitot tube that measures a reference velocity (calibration in yaw was not performed). In particular, the calibration velocities were decomposed into the longitudinal and transversal velocity components by adopting a yaw correction with constant coefficients $K_1^2=K_2^2=0.0225$ \cite{jorgensen2002measure}. The experimental error at a fixed reference location was approximately $\pm 2\%$ for the mean and the standard deviation.
	
	\begin{figure*}[h]
		\centering
		\includegraphics[width=.95\linewidth]{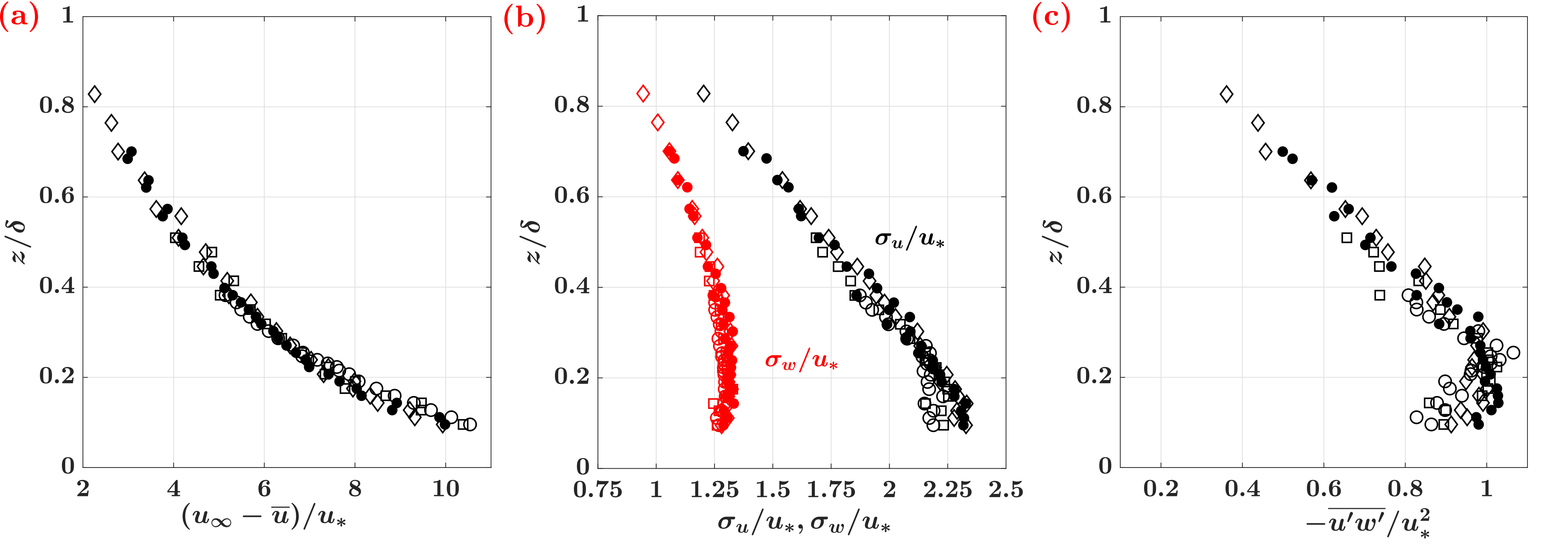}
		\caption{Vertical profiles of velocity statistics normalized with the friction velocity, $u_*$. (a) Mean velocity defect, ${\left(u_\infty - \overline{u}\right)}$. (b) Standard deviations of the longitudinal, ${\sigma_u}$, and vertical, ${\sigma_w}$, velocity component, depicted in black and red, respectively. (c) Reynolds stress, ${-\overline{u'w'}}$. Symbols: ${x/\delta=0.65\quad\circ}$; ${x/\delta=1.30\quad\square}$; ${x/\delta=2.60\quad\bullet}$; ${x/\delta=3.90\quad\lozenge}$.}
		\label{fig:vel_charac}
	\end{figure*}

	A fast Flame Ionization Detector (FID) was used to perform concentration measurements. The FID system uses a sampling tube that is $0.3$ m long, permitting a frequency response of the instrument to about $400$ Hz. The calibration was carried out twice a day by setting ethane-air concentrations equal to 0, 500, 1000 and 5000 ppm. A linear relation holds between ethane concentration and tension response, with slope variations (i.e., the sensitivity variations of the instrument) of about $\pm3\%$, depending on the ambient conditions. The error in the first four moments of the concentration due to all the uncertainties in the experimental chain, was estimated to be up to $4.5\%$. By exploiting measurements performed on different days of distant weeks, the first two moments of the concentration are affected by an error of $2\%$ in the far field and $3\%$ in the near field (this increase is due to uncertainties in the source flow control system in the near field). For the third and fourth moments of the concentration, the error rises up to $4.5\%$ both in the near- and far-field. 
	
	In order to evaluate one-point turbulent fluxes, simultaneous measurements of velocity and concentration are necessary at the same spatial location, implying that HWA and FID systems have to be synchronized and sufficiently close in space. The optimal distance between the HWA and FID was found to be 5 mm in the spanwise direction, in order to avoid local perturbations induced by the measuring system in the flow field. The coupling HWA-FID does not require signal re-sampling or filtering, because both HWA and FID have a constant sampling frequency (equal to 1000 Hz), so that their responses are continuous and regular.

	The main features of the velocity field are shown in Fig. \ref{fig:vel_charac}. Specifically, the vertical profile of the mean velocity defect is displayed in Fig. \ref{fig:vel_charac}(a), while standard deviations of the velocity components, $\sigma_u$ and $\sigma_w$, and the Reynolds stress, $\overline{u'w'}$, are shown in Fig. \ref{fig:vel_charac}(b) and Fig. \ref{fig:vel_charac}(c), respectively. The vertical profiles -- measured at different streamwise locations in the range $x/\delta\in\left[0.65,3.90\right]$ -- are in good agreement with values reported in literature \cite{nironi2015dispersion}, and collapse rather well both for the mean flow (Fig. \ref{fig:vel_charac}(a)) and the velocity fluctuations (Fig. \ref{fig:vel_charac}(b-c)). Specifically, from the Reynolds stress profile we are able to estimate the friction velocity as $u_*=\left(-\overline{u'w'}\right)^{0.5}=0.209$ $\textrm{m/s}$, by averaging the profiles of $\overline{u'w'}$ in the region close to the wall \cite{nironi2015dispersion}.

\section{Plume axis estimation}\label{app:PAE}

	In this section, the vertical position, $h_s^*$, of the actual axis of the plume is estimated for both D3 and D6 as the $z$ coordinate of maximum $\overline{c}(z)$ value. In fact, the maximum value of the mean concentration, $h_s^*$, along $z/\delta$ is not exactly at $z=h_s$, but it decreases downstream from the source mainly due to two factors: the effect of the mean shear $\partial\overline{u}/\partial z$ (where $\overline{u}$ is mean streamwise velocity), and the source wake. 
	
	\begin{figure*}[h]
		\centering
		\includegraphics[width=.97\linewidth]{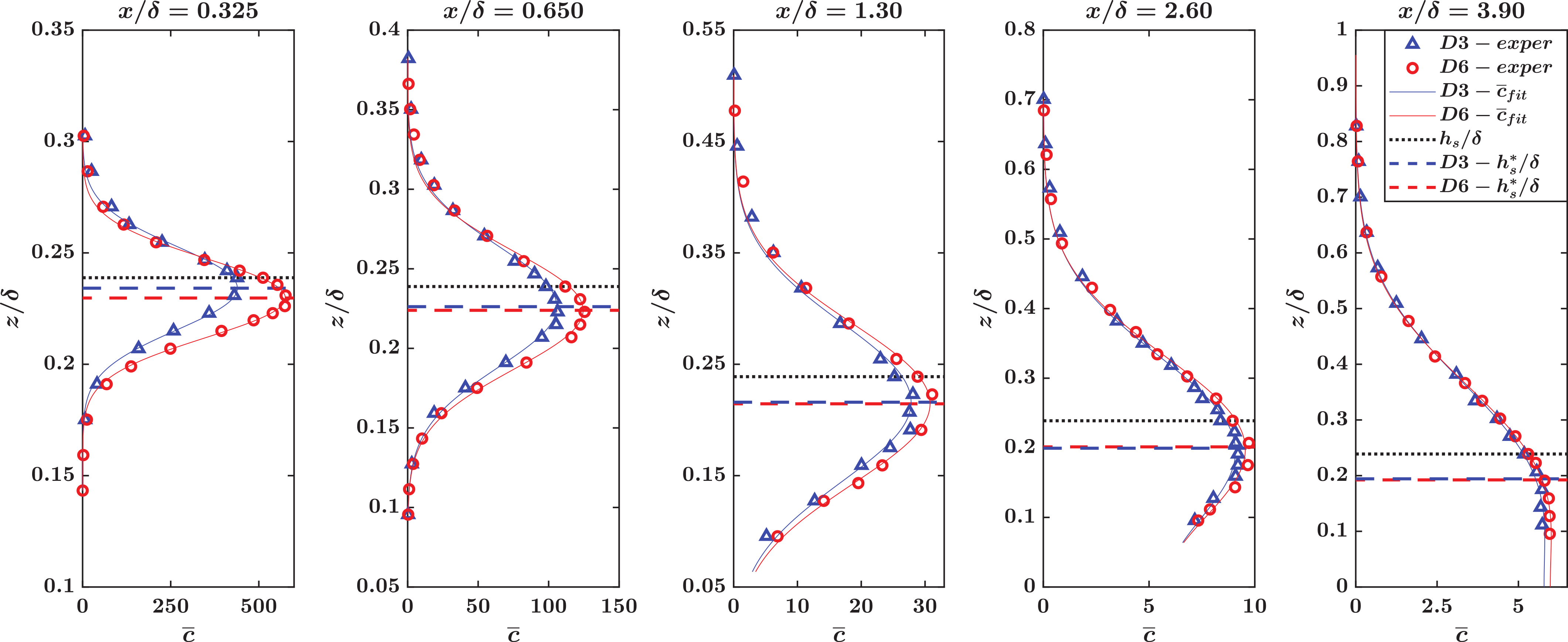}
		\caption{Vertical profiles of the mean concentration value, $\overline{c}$, at different streamwise locations, for the two source diameters, $D=3$ mm and $D=6$ mm. The profiles obtained from the reflected Gaussian distribution of Eq. (\ref{eq:cm_fit}) are also shown. The wall-normal coordinate of the source axis, $h_s$, is illustrated as a horizontal dotted line, while the plume axis height, $h_s^*$, is displayed as a blue (red) dashed line for the source D3 (D6).}\label{figApp:mean_c}
	\end{figure*}	
	
	\noindent Fig. \ref{figApp:mean_c} shows the vertical profiles of the mean concentration, $\overline{c}$, for the two source configurations, D3 and D6, at different streamwise locations, $x/\delta$. In order to extract a reliable value of the wall-normal coordinate of maximum $\overline{c}$, we fitted the vertical profiles of mean concentration by adopting a Gaussian distribution with total reflection on the ground \cite{arya1999air}, defined as

		\begin{equation}\label{eq:cm_fit}
			\overline{c}_{fit}\left(x,z\right)=\frac{M_e/\rho}{2\pi \sigma_y \sigma_z \overline{u}_{adv}}\left[\textrm{exp}\left(-\frac{\left(z+h_s^*\right)^2}{2\sigma_z^2}\right)+ \textrm{exp}\left(-\frac{\left(z-h_s^*\right)^2}{2\sigma_z^2}\right)\right]
		\end{equation}

	\noindent where $\overline{u}_{adv}$ is the mean streamwise velocity at the plume center of mass, $h_s^*$ is the vertical coordinate of the plume axis, while $\sigma_y$ and $\sigma_z$ are the transversal and wall-normal (average) spread of the plume, respectively. The fitting function reported in Eq. (\ref{eq:cm_fit}) is the most suited distribution to reproduce the vertical mean concentration profiles \cite{arya1999air}. In particular, here $\sigma_y$, $\sigma_z$ and $h_s^*$ are adopted as free parameters of the fitting procedure. Although the value of $h_s^*$ can also be set as constant (as an approximation) and equal to $h_s$, in this work we explicitly used $h_s^*$ as a free parameter in the Eq. (\ref{eq:cm_fit}) to highlight the effect of the mean shear on the plume. By doing so, $h_s^*$ is not fixed but depends on the source size by moving downstream, with $h_s^*\leq h_s$. In Fig. \ref{figApp:mean_c}, the values of $h_s^*$ are shown as horizontal dashed and dot-dashed lines for D3 and D6, respectively. It is also worth noting that the values of the vertical coordinate of the plume axis, $h_s^*$, are better identified by the parameters of the fitting procedure than by locating the maximum (experimental) $\overline{c}$ value. This issue is crucial in the far field (i.e., far downstream from the source), where $h_s^*>0$ whereas the maximum (experimental) $\overline{c}$ value is located very close to the floor (i.e., $(z/\delta)_{\overline{c}_{max}}\rightarrow0$, due to the plume dispersion and the effect of the ground reflection of the plume).

	\section{Peak-pit asymmetry in turbulent transport signals}	\label{app:PPATTS}
		
			In this Section, we investigate the asymmetry between peaks and pits in the temporal structure of vertical turbulent transport time-series. This analysis is motivated by the fact that the passive scalar can be transported by turbulence upwards ($w'>0$) and downwards ($w'<0$).
			
			\begin{figure*}[h]
				\centering
				\includegraphics[width=\linewidth]{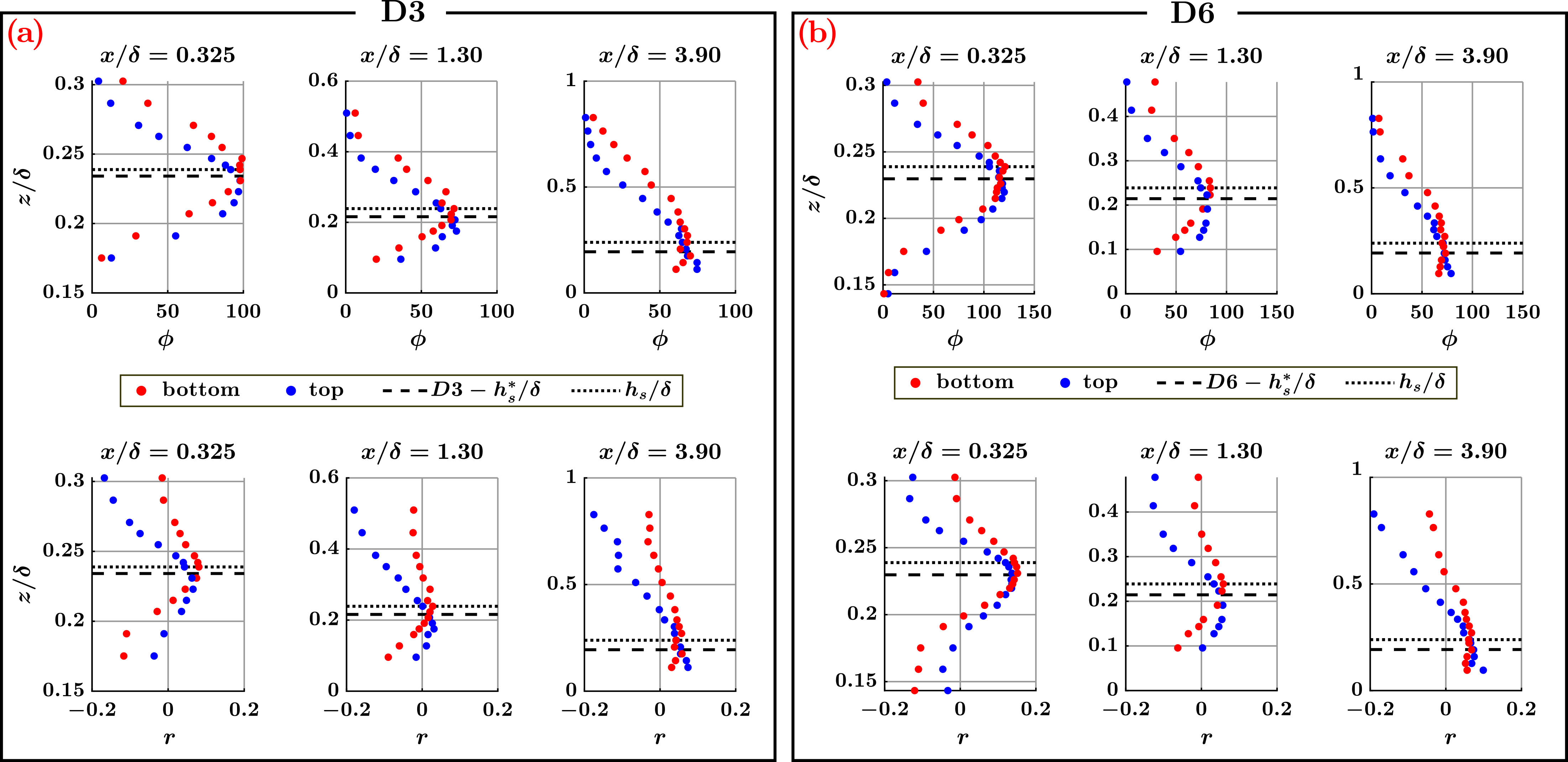}
				\caption{Vertical profiles of the average peak occurrence, $\phi$, and the assortativity coefficient, $r$, for the top- and bottom-visibility. The profiles are plotted for D3 (a) and D6 (b), in the near field ($x/\delta=0.325$), at an intermediate location ($x/\delta=1.30$), and in the far field ($x/\delta=3.90$). The wall-normal coordinate of the source axis, $h_s$, is illustrated as a horizontal dotted line, while the plume axis height, $h_s^*$, is displayed as a dashed horizontal line for both D3 and D6.}\label{fig:metriche_top_bot_wc}
			\end{figure*}

			\noindent To address this issue from the network perspective, in Fig. \ref{fig:metriche_top_bot_wc} we show the behaviour of $\phi$ and $r$ obtained from the series $w'c'$ (top-visibility) and from the series $-w'c'$ (bottom-visibility). The intersection between the metrics from the top- and bottom-visibility accurately corresponds to the vertical location of the actual plume axis, $h_s^*$, for both the source configurations D3 and D6. In fact, in the plume axis the vertical turbulent transport is expected to be symmetrical upwards and downwards (e.g., see Fig. \ref{fig:serie_wc_D3D6}(b,e)). On the other hand, the peak-pit asymmetry is significantly observed in the metric values above and below the plume axis, $h_s^*$. In particular, the metrics from the bottom-visibility are always smaller than the metrics from the top-visibility when the region below the plume axis is focused, while the opposite behaviour is found above the plume axis. The assortativity coefficient, $r$, for the bottom-visibility does not reach strong negative values for $z>h_s^*$, in both cases D3 and D6. This implies that there is not a strong vertical separation between pits and the other data values in the series of $-w'c'$, i.e. extreme negative values are unlikely to appear for $z>h_s^*$ (as shown in Fig. \ref{fig:serie_wc_D3D6}(a,d)). In fact, since most of the passive scalar is present around the plume axis, for $z>h_s^*$ vertical turbulent transport is mainly upwards while for $z<h_s^*$ it is mainly downwards. In order to understand why the average peak occurrence, $\phi$, for the bottom-visibility is larger than for the top-visibility for $z>h_s^*$, we recall that the visibility algorithm is insensitive to absolute intensity of the data in the series. Therefore, above the plume axis, the visibility algorithm mostly considers pits in the series as negative values that are close to zero, because large negative values are very unlikely to appear. As a result, pits are found to frequently appear in time (i.e., high $\phi$) and without a strong vertical separation (i.e., high $r$). On the contrary, the behaviour of the network metrics below the plume axis is the opposite with respect to the metrics above the plume axis, as the series above the plume axis show an opposite temporal structure than the series below the plume axis (e.g., see Fig. \ref{fig:serie_wc_D3D6}(a,d) and Fig. \ref{fig:serie_wc_D3D6}(c,f)).

\bibliography{Iacobello_PRFl_biblio}

\end{document}